\documentclass[aps,pra,twocolumn,showpacs,superscriptaddress,10pt]{revtex4-1}

\usepackage{xcolor, graphicx}
\usepackage{amsmath, amssymb}
\usepackage{ulem}
\usepackage[colorlinks=true,urlcolor=blue,citecolor=blue,linkcolor=blue]{hyperref}

\begin{document}

\title{Operational definition of quantum correlations of light}

\author{J. Sperling}\email{jan.sperling@uni-rostock.de}\affiliation{Arbeitsgruppe Theoretische Quantenoptik, Institut f\"ur Physik, Universit\"at Rostock, D-18051 Rostock, Germany}
\author{W. Vogel}\affiliation{Arbeitsgruppe Theoretische Quantenoptik, Institut f\"ur Physik, Universit\"at Rostock, D-18051 Rostock, Germany}
\author{G. S. Agarwal}\affiliation{Department of Physics, Oklahoma State University, Stillwater, Oklahoma 74078, USA}

\date{\today}

\begin{abstract}
	Quantum features of correlated optical modes define a major aspect of the nonclassicality in quantized radiation fields.
	However, the phase-sensitive detection of a two-mode light field is restricted to interferometric setups and local intensity measurements.
	Even the full reconstruction of the quantum state of a single radiation mode relies on such detection layouts and the preparation of a well-defined reference light field.
	In this work, we establish the notion of the essential quantum correlations of two-mode light fields.
	It refers to those quantum correlations which are measurable by a given device, i.e., the accessible part of a nonclassical Glauber-Sudarshan phase-space distribution, which does not depend on a global phase.
	Assuming a simple four-port interferometer and photon-number-resolving detectors, we derive the reconstruction method and nonclassicality criteria based on the Laplace-transformed moment-generating function of the essential quasiprobability.
	With this technique, we demonstrate that the essential quantum correlations of a polarization tomography scheme are observable even if the detectors are imperfect and cannot truly resolve the photon statistics.
\end{abstract}

\maketitle


\section{Introduction}
	Measuring quantum correlations in optical systems is a key aspect of the vast field of quantum optics~\cite{MW95,VW06,A13}.
	As quantum effects may exhibit a variety of different observable signatures, it is a fundamental, yet cumbersome task to identify them.
	In particular in optical systems, quantum and classical optical interferences may occur simultaneously.
	However, modern applications of quantum light~\cite{BL05,GT07} require proper techniques to discern the domain of classical optics from the features of quantized radiation fields on a measurable basis.

	A well established definition of nonclassical correlations is based on the theory of classical coherence and its violation in quantum systems; see Refs.~\cite{HT56,G63} for early studies.
	For instance, the prominent photon antibunching can be uncovered in this way by measuring intensity correlation functions~\cite{KDM77}.
	For a proper visualization of quantum effects and relating them to a classical frame, quantum-optical phase-space distributions have gained major importance.
	Among the various forms of such quasiprobabilities, the Glauber-Sudarshan representation is the most fundamental one~\cite{G63a,S63}.
	This is due to the fact that nonclassical light is defined as the inability of the interpretation of this particular distribution in terms of classical probability theory for the radiation field under study~\cite{TG65,M86}.

	In order to access the quantum characteristics of a single or multiple optical modes, one can follow two paths.
	On the one hand, one can formulate observable nonclassicality criteria, which may uncover certain quantum effects.
	However, they do not allow for a full identification of nonclassicality or they require an infinite number of tests~\cite{AT92,A93,SRV05,SV05a}.
	Examples of such nonclassicality probes are variance, covariance, or, in general, higher-order moment-based criteria; see Refs.~\cite{RL09,MBWLF10} for overviews.
	On the other hand, a reconstruction of the full quantum state of light is another approach.
	This renders it possible to detect all quantum features in a system.
	However, it requires involved measurement schemes or costly data analysis methods to reconstruct the experimentally realized state~\cite{WVO99,LR09}.
	Examples of such state representations are the reconstruction of the Fock density matrix, quasiprobabilities in phase space, or the characteristic function.
	The latter is the Fourier transform of a phase-space function, it is measurable with balanced homodyne detection, and it can probe the nonclassicality~\cite{V00,LS02,RV02,ZPB07,MKMPE11}.

	Another flaw, which has to be considered, is that these techniques typically require the generation of a proper reference signal or the desired measurements are not available.
	For example, even in classical optics the phase of a signal cannot be directly measured.
	It has to be inferred from an interference with a properly generated reference signal.
	Thus, one can state that not all aspects of the definition of nonclassicality are accessible.
	For this reason, the question of which quantum correlations are truly measurable and not just a mathematical definition is an urgent problem which has to be resolved.
	An operational approach to address this task is the main topic of this contribution.

	In this work, we study the nonclassicality of a bipartite radiation field which is accessible within four-port interferometers, i.e., two input and two output ports, and using local intensity measurements only.
	The family of quantum correlations that are detectable in this manner will define the notion of {\it essential quantum correlations}.
	Using the Schwinger representation of a bipartite system of harmonic oscillators, we derive the corresponding phase-space distribution.
	In a next step, we formulate a method for reconstructing this phase-space function, which is based on the direct measurement of a generalization of the moment-generating function.
	In addition, a hierarchy of nonclassicality criteria is deduced which is formulated in terms of this moment-generating function and the practicability of this technique is studied.
	Finally, we outline an implementation of our approach which is based on state-of-the-art measurement layouts and imperfect detectors that consists of an array of on-off diodes only.

	The article is structured as follows.
	In Sec.~\ref{sec:1}, we derive the operational notion of essential quantum correlations.
	Based on the Laplace transform, a reconstruction technique is elaborated in Sec.~\ref{sec:2}.
	Nonclassicality tests, given by a matrix of the moment-generating function, are formulated in Sec.~\ref{sec:3}.
	A first application is given in Sec.~\ref{sec:HOM}.
	In Sec.~\ref{sec:4}, the theory of an experimental implementation is studied.
	We give a summary and conclude with an outlook in Sec.~\ref{sec:5}.


\section{Essential quantum correlations}\label{sec:1}
	For establishing the essential quantum correlations, let us formulate an operational definition via a typical measurement scheme.
	A two-mode quantum state of light $\hat\rho$ can be expanded in a coherent state basis by applying the Glauber-Sudarshan $P$ representation~\cite{G63a,S63},
	\begin{align}
		\hat \rho=\int_{\mathbb C^2} d^2\alpha d^2\beta P(\alpha,\beta) |\alpha,\beta\rangle\langle\alpha,\beta|.
	\end{align}
	For instance, one could think of a signal field in the first mode and a coherent reference field in the second mode.
	However, this is not required for our approach.
	The two radiation modes may be mixed on a beam splitter, which is described by the input-output relation
	\begin{align}\label{eq:in-out-4port}
		\begin{pmatrix}\hat a_{\boldsymbol e}\\\hat b_{\boldsymbol e}\end{pmatrix}
		=\begin{pmatrix}T&R\\-R^\ast&T^\ast\end{pmatrix}
		\begin{pmatrix}\hat a\\\hat b\end{pmatrix},
	\end{align}
	with $\hat a$ and $\hat b$ ($\hat a_{\boldsymbol e}$ and $\hat b_{\boldsymbol e}$) being the annihilation operators of the input (output) modes and $|T|^2+|R|^2=1$.
	Thus, the output of this four-port interferometer is the transformed state
	\begin{align}
		\hat\rho_{\boldsymbol e}=&\int_{\mathbb C^2} d^2\alpha d^2\beta P(\alpha,\beta)
		\\&\times\nonumber
		|T\alpha+R\beta,T^\ast\beta-R^\ast\alpha\rangle\langle T\alpha+R\beta,T^\ast\beta-R^\ast\alpha|.
	\end{align}
	Further, the intensity of each output mode is detected.
	The result of this measurement can always be written in terms of normally ordered expectation values of operators $\hat F = \hat F(\hat a^\dagger_{\boldsymbol e}\hat a_{\boldsymbol e},
	\hat b^\dagger_{\boldsymbol e}\hat b_{\boldsymbol e})$, which have the general form
	\begin{align}\label{eq:P-Expectation}
		\langle{:}\hat F{:}\rangle{=}\int_{\mathbb C^2} d^2\alpha d^2\beta P(\alpha,\beta) F\left(|T\alpha{+}R\beta|^2,|T^\ast\beta{-}R^\ast\alpha|^2\right)
	\end{align}
	and, naturally, only depend on the output intensities.

	The output mean photon numbers of the initial coherent amplitudes $\alpha$ and $\beta$ can be put in the form
	\begin{align}\label{eq:BosonStokesRelation}
		\frac{\|\boldsymbol S\|\pm\boldsymbol e\cdot\boldsymbol S}{2}=\left\lbrace\begin{array}{ll}
			|T\alpha+R\beta|^2 & \text{for ``$+$''},\\
			|T^\ast\beta-R^\ast\alpha|^2 & \text{for ``$-$''},
		\end{array}\right.
	\end{align}
	where we introduce the real-valued, three-dimensional vectors
	\begin{align}\label{eq:Def-e-and-S}
		\boldsymbol e=\begin{pmatrix}
			2{\rm Re}(TR^\ast)\\2{\rm Im}(TR^\ast)\\|T|^2-|R|^2
		\end{pmatrix}
		\text{ and }
		\boldsymbol S=\begin{pmatrix}
			2{\rm Re}(\alpha^\ast\beta)\\2{\rm Im}(\alpha^\ast\beta)\\|\alpha|^2-|\beta|^2
		\end{pmatrix},
	\end{align}
	with $\boldsymbol e\cdot\boldsymbol S=e_xS_x+e_yS_y+e_zS_z$ being the standard scalar product and its corresponding norm
	\begin{align}
		\|\boldsymbol S\|=\sqrt{\boldsymbol S\cdot\boldsymbol S}.
	\end{align}
	It is worth mentioning that the defined vector $\boldsymbol e$ also justifies the index of the output operators, e.g., in Eq.~\eqref{eq:in-out-4port}, and that $\boldsymbol e$ is normalized, $\|\boldsymbol e\|=1$, as we have $|T|^2+|R|^2=1$.
	The initial coherent amplitudes are retrieved via the inverse transformation,
	\begin{align}\label{eq:transfo}
		\begin{pmatrix}\alpha\\\beta\end{pmatrix}=\begin{pmatrix}
			\sqrt{\frac{\|\boldsymbol S\|+S_z}{2}}\exp\left[i\frac{\sigma-\arg(S_x+iS_y)}{2}\right]
			\\ \sqrt{\frac{\|\boldsymbol S\|-S_z}{2}}\exp\left[i\frac{\sigma+\arg(S_x+iS_y)}{2}\right]
		\end{pmatrix},
	\end{align}
	where $\sigma$ denotes a non-specified, global phase, the relative phase is $\arg(S_x+iS_y)=\arg(\alpha^\ast\beta)$, and we have $\|\boldsymbol S\|=|\alpha|^2+|\beta|^2$.

	Now, the expectation value of the four-port interferometer in Eq.~\eqref{eq:P-Expectation} takes the following form:
	\begin{align}
		\langle{:}\hat F{:}\rangle=\int_{\mathbb R^3} d^3\boldsymbol S P_{\rm ess}(\boldsymbol S)F\left(\frac{\|\boldsymbol S\|+\boldsymbol e\cdot\boldsymbol S}{2},\frac{\|\boldsymbol S\|-\boldsymbol e\cdot\boldsymbol S}{2}\right),
	\end{align}
	with the {\it essential} Glauber-Sudarshan distribution $P_{\rm ess}$,
	\begin{align}
		P_{\rm ess}(\boldsymbol S)=\frac{1}{8\|\boldsymbol S\|}\int_0^{2\pi} d\sigma P(\alpha,\beta).
	\end{align}
	It has been obtained via the above introduced coordinate transformation $(\boldsymbol S,\sigma)\mapsto (\alpha,\beta)$ in Eq.~\eqref{eq:transfo}.

	In the here-presented operational sense, the notation essential shall indicate that $P_{\rm ess}$ includes the full information that is accessible with the detection scenario under consideration.
	The distribution $P_{\rm ess}$ depends on three real-valued parameters, $S_x$, $S_y$, and $S_z$, rather than the two complex parameters $\alpha$ and $\beta$, which is the result of the averaging over the non-accessible global phase $\sigma$.
	Moreover, in this operational meaning, the state is {\it essentially classical} if $P_{\rm ess}$ is a positive semidefinite distribution.
	Otherwise, we have an essentially quantum-correlated light field.

	In case one restricts to polarization measurements, the physical meaning of the vector $\boldsymbol S$ is related to the Stokes parameters which characterize the various forms of the polarization of a light beam~\cite{LS00,BSSKGMM10}.
	Here, $\boldsymbol S$ applies to any two degrees of freedom, which relates to the Schwinger representation of two harmonic oscillators; see Ref.~\cite{CMMSZ06} for a review.
	We have
	\begin{align}\label{eq:DefStokes}
		\boldsymbol S=\langle \alpha,\beta|\boldsymbol{\hat S}|\alpha,\beta\rangle,
		\text{ with }
		\boldsymbol{\hat S}=\begin{pmatrix}
			\hat a^\dag\hat b+\hat b^\dag\hat a\\
			-i\hat a^\dag\hat b+i\hat b^\dag\hat a\\
			\hat a^\dag\hat a-\hat b^\dag\hat b\\
		\end{pmatrix},
	\end{align}
	cf. Eq.~\eqref{eq:Def-e-and-S}.
	In addition, the norm of $\boldsymbol S$ is obtained through the total photon number,
	\begin{align}\label{eq:DefTotalNumber}
		\|\boldsymbol S\|=\langle\alpha,\beta|\hat N|\alpha,\beta\rangle
		\text{ for $\hat N=\hat a^\dag\hat a+\hat b^\dag\hat b$},
	\end{align}
	which is also referred to as the zeroth component of the Stokes vector in polarization measurements; $S_0=\|\boldsymbol S\|$, likewise $\hat S_0=\hat N$.
	Also note that the total photon number is invariant under beam splitter transformations, i.e., $\hat N=\hat a^\dag\hat a+\hat b^\dag\hat b=\hat a_{\boldsymbol e}^\dag\hat a_{\boldsymbol e}+\hat b_{\boldsymbol e}^\dag\hat b_{\boldsymbol e}$.
	The operator representation of relation~\eqref{eq:BosonStokesRelation} is
	\begin{align}\label{eq:output-numbers}
		\frac{\hat N+\boldsymbol e\cdot\boldsymbol{\hat S}}{2}=\hat a_{\boldsymbol e}^\dag\hat a_{\boldsymbol e}
		\text{ and }
		\frac{\hat N-\boldsymbol e\cdot\boldsymbol{\hat S}}{2}=\hat b_{\boldsymbol e}^\dag\hat b_{\boldsymbol e}.
	\end{align}
	Let us also mention that the Stokes operators have the properties of a spin angular momentum algebra~\cite{LS00,BSSKGMM10,CMMSZ06}, with $\boldsymbol{\hat J}=\boldsymbol{\hat S}/2$ and $\hat J^2=\boldsymbol{\hat J}\cdot\boldsymbol{\hat J}=(\hat N/2)(\hat N/2+\hat 1)$.

	As we restrict ourselves to linear scenarios, let us also briefly comment on other types of interference schemes.
	Our approach could be also formulated by using nonlinear interferometers.
	For instance, they could be based on parametric processes~\cite{YMK86,AB01,JLZOZ11}.
	These $\mathrm{SU}(1,1)$ interferometers have been applied, for example, in quantum metrology~\cite{PDA10,HKLJOZ14}.


\section{Reconstruction of $P_{\rm ess}$ and the moment-generating function}\label{sec:2}
	In order to reconstruct the essential phase-space distribution, we make use of the moment-generating function (MGF).
	In contrast to the typically considered approach~\cite{VW06,A13}, which applies the characteristic function (the Fourier transform of the $P$ distribution), we apply the MGF (the Laplace transform of $P$).
	Therefore, let us briefly recall some properties of the MGF in a one-dimensional setting.

	The MGF of a distribution $P(x)$ is defined as the expectation value of $\exp(tx)$,
	\begin{align}
		M(t)=\int_{\mathbb R} dx P(x)e^{tx}.
	\end{align}
	In this form, $M(-t)$ corresponds to the two-sided Laplace transform of $P(x)$~\cite{W15}.
	The derivatives of $M(t)$ yield the moments of $P(x)$.
	In addition, $M(ik)=\Phi(k)$ is the characteristic function of $P(x)$.
	That is, the MGF for purely imaginary arguments yields the Fourier transform of $P(x)$.
	From this relation, the well-known, inverse two-sided Laplace transformed was introduced,
	\begin{align}
		P(x)=\int_{i\mathbb R} \frac{dt}{2\pi i} e^{-tx} M(t),
	\end{align}
	which serves as our reconstruction approach.

	Let us now consider the three-dimensional MGF in the context of our phase-space distribution $P_{\rm ess}(\boldsymbol S)$.
	Namely, we have the identities
	\begin{align}\label{eq:Pess-and-Phiess}
	\begin{aligned}
		M_{\rm ess}(\boldsymbol t;\tau)=&\int_{\mathbb R^3} d^3\boldsymbol S P_{\rm ess}(\boldsymbol S) e^{\boldsymbol t\cdot\boldsymbol S-\tau\|\boldsymbol S\|}
		\\\text{and }
		P_{\rm ess}(\boldsymbol S)=&\frac{e^{\tau\|\boldsymbol S\|}}{(2\pi i)^3}\int_{i\mathbb R^3} d^3\boldsymbol t M_{\rm ess}(\boldsymbol t;\tau) e^{-\boldsymbol t\cdot\boldsymbol S},
	\end{aligned}
	\end{align}
	where we additionally introduce the converging factor $\tau>0$.
	It guarantees the existence of $M_{\rm ess}(\boldsymbol t;\tau)$ which is assured for any $\|{\rm Re}(\boldsymbol t)\|\leq \tau$, since the integral kernel in Eq.~\eqref{eq:Pess-and-Phiess} is upper bounded by 1 in this case~\cite{W15}, $|\exp(\boldsymbol t\cdot\boldsymbol S-\tau\|\boldsymbol S\|)|\leq 1$.
	From this definition of the {\it essential} MGF, $M_{\rm ess}(\boldsymbol t;\tau)$, we can see that it fulfills the normalization relation, $M_{\rm ess}(0;0)=1$, and the symmetry relation, $M_{\rm ess}(\boldsymbol t^\ast;\tau)=M_{\rm ess}(\boldsymbol t;\tau)^\ast$.

	Applying definitions~\eqref{eq:DefStokes} and~\eqref{eq:DefTotalNumber}, as well as the coordinate transformation~\eqref{eq:transfo}, we obtain
	\begin{align}
		&M_{\rm ess}(\boldsymbol t;\tau)	\nonumber
		\\=&\int_{\mathbb C^2} d^2\alpha d^2\beta P(\alpha,\beta)
		\langle\alpha,\beta|{:}
			\exp\left[t\boldsymbol e\cdot\boldsymbol{\hat S}-\tau\hat N\right]
		{:}|\alpha,\beta\rangle\nonumber
		\\=&\langle{:}\exp\left[t \boldsymbol e\cdot\boldsymbol{\hat S}-\tau\hat N\right]{:}\rangle,
		\label{eq:expectMGF0}
	\end{align}
	where we decompose $\boldsymbol t=t\boldsymbol e$.
	Thus, certain values of $T$ and $R$, which describe the beam splitter in Eq.~\eqref{eq:in-out-4port} and the vector $\boldsymbol e$ in Eq.~\eqref{eq:Def-e-and-S}, define different directions in the Laplace-transformed, essential phase space which is represented by $M_{\rm ess}(t\boldsymbol e;\tau)$.
	Employing Eq.~\eqref{eq:output-numbers}, we can further write
	\begin{align}\nonumber
		&M_{\rm ess}(t\boldsymbol e;\tau)=\langle{:}\exp\left[(t-\tau)\hat a_{\boldsymbol e}^\dag\hat a_{\boldsymbol e}+(-t-\tau)\hat b_{\boldsymbol e}^\dag\hat b_{\boldsymbol e}\right]{:}\rangle
		\\=&\left\langle(1+t-\tau)^{\hat a_{\boldsymbol e}^\dag\hat a_{\boldsymbol e}}(1-t-\tau)^{\hat b_{\boldsymbol e}^\dag\hat b_{\boldsymbol e}}\right\rangle,
		\label{eq:expectMGF}
	\end{align}
	where ${:}\exp[x\hat n]{:}=(1-x)^{\hat n}$ is used for the corresponding photon-number operator $\hat n$~\cite{AW70}.
	Hence, the MGF can be directly obtained by measuring the joint photon-number distribution $p(n_a,n_b;\boldsymbol e)$ of the two modes,
	\begin{align}\label{eq:sampleMess}
	\begin{aligned}
		&M_{\rm ess}(t\boldsymbol e;\tau)
		\\=&\sum_{n_a,n_b=0}^\infty (1+t-\tau)^{n_a}(1-t-\tau)^{n_b} p(n_a,n_b;\boldsymbol e).
	\end{aligned}
	\end{align}
	Therefore, we have shown that the essential MGF $M_{\rm ess}$ is directly measurable.
	Let us emphasize that $M_{\rm ess}(t\boldsymbol e;\tau)$ will exist unless the geometric series in Eq.~\eqref{eq:sampleMess} diverges which is unlikely to happen for most physical scenarios.
	Additionally, we will generalize this approach to include imperfect detection scenarios in Sec.~\ref{sec:4}.

	It is also worth mentioning that for most applications, the number of photons is finite ($n_a+n_b\leq N$) and, thus, the sum in Eq.~\eqref{eq:sampleMess} is a finite one as well.
	In this case, $M_{\rm ess}(t\boldsymbol e;\tau)$ represents a polynomial of the two variables $t$ and $\tau$ of the maximal degree $N$ for any $\boldsymbol e$.
	For a fixed $\tau$ value and a real-valued $t$, the MGF can have a root.
	Conversely, any classical mixture of coherent states is described by strictly positive MGF in that case, $M_{\rm ess}(t\boldsymbol e;\tau)>0$ [see Eq.~\eqref{eq:expectMGF0}].
	The nonclassical feature,
	\begin{align}\label{eq:node}
		M_{\rm ess}(t\boldsymbol e;\tau)=0,
	\end{align}
	is also discussed in Sec.~\ref{sec:HOM} for an example.

	Let us also discuss the MGF in relation to the characteristic function.
	As pointed out earlier, for imaginary arguments $i\boldsymbol k=\boldsymbol t$ with $\boldsymbol k\in\mathbb R^3$, we have the essential characteristic function,
	\begin{align}\label{eq:CFess}
		\Phi_{\rm ess}(\boldsymbol k)=M_{\rm ess}(i\boldsymbol k;0)=\langle{:}\exp[i\boldsymbol k\cdot\boldsymbol{\hat S}]{:}\rangle,
	\end{align}
	for $\tau=0$.
	For comparison, the two-mode characteristic function of the Glauber-Sudarshan representation is $\Phi(\alpha',\beta')=\langle{:}\exp[\alpha'\hat a^\dag-\alpha'{}^\ast\hat a+\beta'\hat b^\dag-\beta'{}^\ast\hat b]{:}\rangle$.
	The characteristic function $\Phi_{\rm ess}(\boldsymbol k)$ in Eq.~\eqref{eq:CFess} depends on the photon-number operators, as we have $\boldsymbol e\cdot\boldsymbol{\hat S}=\hat a_{\boldsymbol e}^\dag\hat a_{\boldsymbol e}-\hat b_{\boldsymbol e}^\dag\hat b_{\boldsymbol e}$.
	By contrast, the typical characteristic function $\Phi(\alpha',\beta')$ is defined in terms of the, in our scenario, not directly measurable field operators $\hat a$ and $\hat b$.

	The inverse transformation in Eq.~\eqref{eq:Pess-and-Phiess} allows one, in principle, to reconstruct the $P_{\rm ess}$ from the directly measurable MGF $M_{\rm ess}$.
	However, the inverse transform might not be possible in real experiments, due to a singular behavior of the Glauber-Sudarshan distribution.
	Thus, let us formulate general nonclassicality criteria directly in terms of the MGF.


\section{Nonclassicality tests via $M_{\rm ess}$}\label{sec:3}
	The formulation of nonclassicality probes has always been a major subject of research in quantum optics; cf. Refs.~\cite{VW06,A13}.
	If the Glauber-Sudarshan distribution is a regular one, one can reconstruct $P$ with negative contributions.
	In case this is impossible, the characteristic function can also yield the nonclassicality, because a number of directly accessible nonclassicality conditions have been formulated on the basis of Bochner's theorem~\cite{V00,RV02}.
	Let us apply a similar technique to infer the essential nonclassicality employing the MGF.

	The fundamental benchmark for nonclassicality is the existence of an operator $\hat f$ such that
	\begin{align}\label{eq:fundNCL}
		0>\langle{:}\hat f^\dag\hat f{:}\rangle.
	\end{align}
	For the aim of a characterization based on the MGF, it is favorable to expand $\hat f=\sum_p f_p \exp[-\tau_p\hat N+\boldsymbol t_p\cdot\boldsymbol{\hat S}]$, which gives
	\begin{align}
		\langle{:}\hat f^\dag\hat f{:}\rangle{=}\sum_{p,q} f_p^\ast f_q\langle{:}
			\exp\left[{-}(\tau_p+\tau_q)\hat N{+}\left(\boldsymbol t_p^\ast+\boldsymbol t_q\right)\cdot\boldsymbol{\hat S}\right]
		{:}\rangle.
	\end{align}
	In combination with inequality~\eqref{eq:fundNCL}, this means that if the matrix $\boldsymbol M_{\rm ess}$ is not a positive semidefinite one, we have nonclassical light,
	\begin{align}\label{eq:genMessMOM}
		0\nleq\boldsymbol M_{\rm ess}=\left(M_{\rm ess}(\boldsymbol t_p^\ast+\boldsymbol t_q;\tau_p+\tau_q)\right)_{p,q\in\mathcal I},
	\end{align}
	for an arbitrary choice of $\tau_r$ and $\boldsymbol t_r$ in an index set $\mathcal I$ ($r\in\mathcal I$).
	Using Silvester's criterion, we get the following: If the determinant of $\boldsymbol M_{\rm ess}$ is negative, $0>\det\boldsymbol M_{\rm ess}$, for a given $\mathcal I$ with an arbitrary cardinality, we have verified the nonclassicality.

	Moreover, let us mention that the choices $\tau_r=0$ and $\boldsymbol t_r\in i\mathbb R^3$ yields the nonclassicality criteria similar to those of the characteristic function in terms of the field operators $\hat a$ and $\hat b$.
	In this case, our family of nonclassicality conditions is directly related to the hierarchy of corresponding criteria in Ref.~\cite{RV02}.
	Therefore, our MGF nonclassicality criteria are also necessary and sufficient ones to infer the essential quantum correlations.

	Due to its physical relevance (see, e.g., Ref.~\cite{MBWLF10}), let us study in more detail the second-order nonclassicality criterion, which reads
	\begin{align}\label{eq:second-order-NCL}
		0>\det\begin{pmatrix}
			M_{\rm ess}(2{\rm Re}[\boldsymbol t];2\tau) & M_{\rm ess}(\boldsymbol t^\ast+\boldsymbol t';\tau+\tau')\\
			M_{\rm ess}(\boldsymbol t^\ast+\boldsymbol t';\tau+\tau')^\ast & M_{\rm ess}(2{\rm Re}[\boldsymbol t'];2\tau')
		\end{pmatrix}.
	\end{align}
	First, if we set $\tau=\tau'=0$ and $\boldsymbol t,\boldsymbol t'\in i\mathbb R^3$, we retrieve the nonclassicality criteria in terms of the characteristic function	~\cite{V00},
	\begin{align}
		|\Phi_{\rm ess}(\boldsymbol k)|>1,
	\end{align}
	with $\Phi_{\rm ess}(\boldsymbol k)=M_{\rm ess}(i\boldsymbol k,0)$ and $\boldsymbol k=i(\boldsymbol t-\boldsymbol t')$.
	Second, the choices $\tau'=0$ and $\boldsymbol t'=0$ yield a nonclassicality condition in terms of a negative, normally ordered variance,
	\begin{align}
		0>\langle{:}\Delta \hat A^\dag\Delta\hat A{:}\rangle,
	\end{align}
	with $\Delta\hat A=\hat A-\langle{:}\hat A{:}\rangle$ and $\hat A=\exp[-\tau\hat N+\boldsymbol t\cdot\boldsymbol{\hat S}]$.
	If we further take $\hat B=\exp[-\tau'\hat N+\boldsymbol t'\cdot\boldsymbol{\hat S}]$ (for $\tau'\neq0$ or $\boldsymbol t'\neq0$), we find that condition~\eqref{eq:second-order-NCL} can be also understood as the violation of a Cauchy-Schwarz inequality~\cite{A88}.
	That is,
	\begin{align}
		|\langle{:} A^\dagger\hat B{:}\rangle|^2>\langle{:} \hat A^\dag\hat A{:}\rangle\langle{:}\hat B^\dag\hat B{:}\rangle
	\end{align}
	verifies essential nonclassical correlations.

	Moreover, low-intensity light fields are studied in many experiments.
	Therefore, we may also consider this limit that allows one to perform a second-order Taylor expansion of the exponential functions in inequality~\eqref{eq:second-order-NCL}.
	In this limit, we have the following for $\boldsymbol t=t\boldsymbol e$ and $\boldsymbol t'=t'\boldsymbol e$:
	\begin{align}
		0>\langle{:}\left[\Delta\left([\tau'-\tau]\hat N-[t'-t]\boldsymbol e\cdot\boldsymbol{\hat S}\right)\right]^2{:}\rangle,
	\end{align}
	with $t,t'\in\mathbb R$ and $\boldsymbol e\in\mathbb R^3$.
	Now, we can also select different parameters for $\tau'-\tau$ and $t'-t$.
	For instance, we can take $\tau=\tau'$ or $t=t'$ to obtain a condition in terms of the normally ordered variance of $\boldsymbol e\cdot\boldsymbol{\hat S}$ or $\hat N$, 
	\begin{align}\label{eq:simplemomentcriteria}
		\langle{:}(\Delta[\boldsymbol e\cdot\boldsymbol{\hat S}])^2{:}\rangle<0
		\text{ or }
		\langle{:}(\Delta\hat N)^2{:}\rangle<0,
	\end{align}
	respectively.
	It is worth pointing out that the Mandel $Q$ parameter~\cite{M79} for the total photon number $\hat N$ is negative if $\langle{:}(\Delta\hat N)^2{:}\rangle<0$.
	If the conditions~\eqref{eq:simplemomentcriteria} fail to identify the nonclassicality, one can also chose the parameters such that we get -- up to a positive scaling factor -- a cross-correlation condition between the total photon number and the phase-space variable $\boldsymbol S$,
	\begin{align}\label{eq:crosscor}
		0>\langle{:}(\Delta\hat N)^2{:}\rangle\langle{:}(\Delta[\boldsymbol e\cdot\boldsymbol{\hat S}])^2{:}\rangle
		-\langle{:}(\Delta\hat N)(\Delta[\boldsymbol e\cdot\boldsymbol{\hat S}]){:}\rangle^2,
	\end{align}
	based on the choice $(t'-t)(\tau'-\tau)\propto\langle{:}\Delta\hat N)(\Delta[\boldsymbol e\cdot\boldsymbol{\hat S}]){:}\rangle$ and adjusting $(t'-t)/(\tau'-\tau)$ correspondingly.
	In the same fashion, we can use relation~\eqref{eq:output-numbers} to get nonclassical correlations between the photon numbers of the individual detectors,
	\begin{align}
	\begin{aligned}
		0>&\langle{:}(\Delta[\hat a_{\boldsymbol e}^\dagger\hat a_{\boldsymbol e}])^2{:}\rangle\langle{:}(\Delta[\hat b_{\boldsymbol e}^\dagger\hat b_{\boldsymbol e}])^2{:}\rangle
		\\&-\langle{:}(\Delta[\hat a_{\boldsymbol e}^\dagger\hat a_{\boldsymbol e}])(\Delta[\hat b_{\boldsymbol e}^\dagger\hat b_{\boldsymbol e}]){:}\rangle^2.
	\end{aligned}
	\end{align}
	Other second-order inequalities that may be obtained for other choices of parameters $t$, $t'$, $\tau$, and $\tau'$ relate to uncertainty relations for angular momentum~\cite{SB16,DSW15} or in atomic systems~\cite{WBIMH92,A13}.

	Even more directly, the nonclassicality criteria in terms of moments of the field operators and the characteristic function was unified in Ref.~\cite{RSAMKHV15} together with an experimental implementation.
	The idea is that the derivatives of the characteristic function yield the moments, which can be sampled from the experimental data.
	This approach can be adapted for the MGF, as derivatives similarly yield the moments.
	Thus, moments of the form $\langle{:}\hat N^k[\boldsymbol e\cdot\boldsymbol{\hat S}]^l{:}\rangle$ are also accessible beyond the above-discussed low-intensity approximation.

	As one can see, a number of known and unknown nonclassicality criteria are obtained already from the second order of our approach.
	For instance, we have shown that correlations between the components of $\boldsymbol{\hat S}$ and the total photon number $\hat N$ are accessible; see Eq.~\eqref{eq:crosscor} and Refs.~\cite{V08,C10} for the related theory of field-intensity correlations and  Refs.~\cite{CCFO00,GRSECB09} for corresponding experiments.
	Including higher-order matrices of MGF $\boldsymbol M_{\rm ess}$, this can be even extended to cover more complex physical scenarios and to capture all forms of the essential nonclassicality.
	In the following, we start with the verification of the essential quantum correlations for one example.
	Eventually, in Sec.~\ref{sec:4}, we describe a technique that renders it possible to access the function $M_{\rm ess}(\boldsymbol t;\tau)$, at least in parts, in realistic measurement scenarios.


\section{The Husimi function and the Hong-Ou-Mandel effect}\label{sec:HOM}
	The practicability of the previously introduced nonclassicality criteria is shown for a typical, nonclassical effect.
	To do so and as the $P(\alpha,\beta)$ function can be highly singular [the same holds for $P_{\rm ess}(\boldsymbol S)$], we relate Eq.~\eqref{eq:expectMGF0} or~\eqref{eq:expectMGF} to another phase-space representation.
	For this reason, let us define
	\begin{align}\label{eq:lambdas}
		\lambda_a=\tau-t\text{ and }\lambda_b=\tau+t,
	\end{align}
	which trivially yields
	\begin{align}
		M_{\rm ess}(t\boldsymbol e;\tau)=\langle{:}\exp\left[
			-\lambda_a\hat a_{\boldsymbol e}^\dag\hat a_{\boldsymbol e}
			-\lambda_b\hat b_{\boldsymbol e}^\dag\hat b_{\boldsymbol e}
		\right]{:}\rangle.
	\end{align}
	Using the relation
	\begin{align}
		{:}\exp[-\lambda\hat n]{:}=\int_{\mathbb C} d^2\gamma \frac{\exp\left[-\frac{\lambda|\gamma|^2}{1-\lambda}\right]}{\pi(1-\lambda)}|\gamma\rangle\langle\gamma|, 
	\end{align}
	cf. Ref.~\cite{AW70}, we can further write
	\begin{align}\nonumber
		M_{\rm ess}(t\boldsymbol e;\tau)=\int_{\mathbb C^2} d^2\alpha_{\boldsymbol e}d^2\beta_{\boldsymbol e}&
		\frac{\exp\left[
			-\frac{\lambda_a|\alpha_{\boldsymbol e}|^2}{1-\lambda_a}
			-\frac{\lambda_b|\beta_{\boldsymbol e}|^2}{1-\lambda_b}
		\right]}{(1-\lambda_a)(1-\lambda_b)}
		\\&\times\frac{\langle\alpha_{\boldsymbol e},\beta_{\boldsymbol e}|\hat\rho|\alpha_{\boldsymbol e},\beta_{\boldsymbol e}\rangle}{\pi^2}.
	\end{align}
	The latter term defines the Husimi $Q$ function~\cite{H40}, which is always non-negative and well behaved.
	Keeping in mind that the unitary input-output relation~\eqref{eq:in-out-4port} can be equivalently written for the coherent states, we can perform an integral transformation and we obtain
	\begin{align}\label{eq:essMHusimi}
		&M_{\rm ess}(t\boldsymbol e;\tau)\\
		=&\int_{\mathbb C^2} d^2\alpha d^2\beta
		\frac{\exp\left[
			{-}\frac{\lambda_a|T\alpha+R\beta|^2}{1-\lambda_a}
			{-}\frac{\lambda_b|T^\ast\beta-R^\ast\alpha|^2}{1-\lambda_b}
		\right]}{(1-\lambda_a)(1-\lambda_b)}
		Q(\alpha,\beta),\nonumber
	\end{align}
	with the two-mode Husimi $Q$ function:
	\begin{align}\label{eq:Husimi}
		Q(\alpha,\beta)=\frac{\langle\alpha,\beta|\hat\rho|\alpha,\beta\rangle}{\pi^2}.
	\end{align}
	This approach is applied for the following examples.
	Moreover, it uncovers a relation between the MGF $M_{\rm ess}$ and another phase-space distribution.

	Besides other phenomena, the Hong-Ou-Mandel interference experiment is one of the most prominent examples of a nonclassical effect~\cite{HOM87}, which is still a cornerstone for modern quantum optics; see, e.g., Refs.~\cite{LIACBW15,JC15} for recent realizations.
	The basic idea behind this signature of quantumness is the mixture of photons at a beam splitter, one at each input.
	This results in finding both photons at one output only, for certain $T$ and $R$ values.

	Let us analyze the nonclassicality with our approach.
	The input and the output states in the Fock basis are
	\begin{align}\label{eq:HOMstate}
		|\psi\rangle=&|1,1\rangle
		\text{ and }\\\nonumber
		|\psi_{\boldsymbol e}\rangle=&\sqrt2TR|2,0\rangle+(|T|^2{-}|R|^2)|1,1\rangle-\sqrt2T^\ast R^\ast |0,2\rangle,
	\end{align}
	respectively.
	Thus, for a balanced beam splitter, $|T|=|R|$, we observe the above-described phenomena.
	The Husimi function~\eqref{eq:Husimi} and the integral~\eqref{eq:essMHusimi} can be straightforwardly computed.
	We find
	\begin{align}\label{eq:essMHOM}
	\begin{aligned}
		M_{\rm ess}(t\boldsymbol e;\tau)=&
		2|T|^2|R|^2\left[(1-\lambda_a)^2+(1-\lambda_b)^2\right]\\
		&+\left[|T|^2-|R|^2\right]^2(1-\lambda_a)(1-\lambda_b).
	\end{aligned}
	\end{align}
	Using Eq.~\eqref{eq:lambdas} together with the components of the normalized vector in Eq.~\eqref{eq:Def-e-and-S}, $\boldsymbol e=(e_x,e_y,e_z)^\mathrm{T}$ and $e_x^2+e_y^2+e_z^2=1$, we can also write the previous equation in the form
	\begin{align}
		M_{\rm ess}(t\boldsymbol e;\tau)=&
		(1-\tau)^2+(1-2e_z^2)t^2.
	\end{align}
	It is worth mentioning that the $P$ function for this state includes second-order derivatives of the Dirac delta distribution and it is therefore not directly measurable.

	In addition and for comparison, the essential MGF for a coherent state $|\alpha,\beta\rangle$ is
	\begin{align}\label{eq:essMcoherent}
	\begin{aligned}
		&M_{\rm ess}(t\boldsymbol e;\tau)=e^{-\lambda_a|T\alpha+R\beta|^2-\lambda_b|T^\ast\beta-R^\ast\alpha|^2}
		\\=&e^{-\tau\|\boldsymbol S\|+\boldsymbol t\cdot\boldsymbol S}=e^{[-\tau+t\cos(\vartheta)]\|\boldsymbol S\|},
	\end{aligned}
	\end{align}
	with $\boldsymbol S$ and $\boldsymbol e$ as given in Eq.~\eqref{eq:Def-e-and-S} and $\vartheta$ is the angle between $\boldsymbol e$ and $\boldsymbol S$.
	Thus, we have $M_{\rm ess}(t\boldsymbol e;\tau)=e^{-(\tau\pm t)\|\boldsymbol S\|}$ for $\cos(\vartheta)=\mp1$ and $M_{\rm ess}(t\boldsymbol e;\tau)=e^{-\tau\|\boldsymbol S\|}$ for $\cos(\vartheta)=0$.
	For a proper visualization of the MGF as a function of the normalized vector $\boldsymbol e$, let us consider the following map:
	\begin{align}\label{eq:surfacemap}
		\boldsymbol e\mapsto M_{\rm ess}(t\boldsymbol e;\tau)\boldsymbol e.
	\end{align}
	This map transforms the unit sphere, given by the normalized arguments $\|\boldsymbol e\|=1$, into a deformed surface.
	For coherent states and real $t$ values, the map~\eqref{eq:surfacemap} yields an ellipsoid with one major principle axis along $\boldsymbol S$ with the length $e^{-\tau\|\boldsymbol S\|}\cosh(t\|\boldsymbol S\|)$ and two minor principle axes perpendicular to $\boldsymbol S$ with the length $e^{-\tau\|\boldsymbol S\|}$.
	Yet, this function can also have a very different shape if the considered state is a nonclassical one, as we discuss now.

\begin{figure}[t]
	\includegraphics[width=5.5cm]{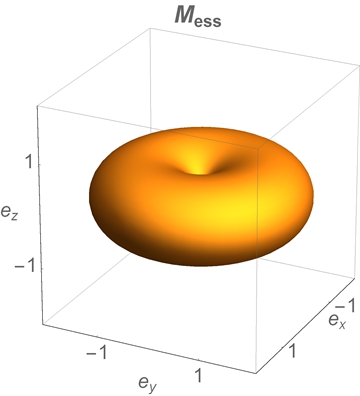}
	\caption{(Color online)
		The plot shows $M_{\rm ess}(\boldsymbol e;0)\boldsymbol e$ [cf. the map in Eq.~\eqref{eq:surfacemap} for $t=1$ and $\tau=0$],
		that is, the essential MGF evaluated on a unit sphere, $\boldsymbol e$, and plotted in the direction of $\boldsymbol e$.
		For the $z$ direction, $\boldsymbol e=(0,0,\pm1)^{\rm T}$, we have $M_{\rm ess}(\boldsymbol e;0)=0$.
	}\label{fig:HOM1}
\end{figure}

	In Fig.~\ref{fig:HOM1}, we show this surface plot for the essential MGF $M_{\rm ess}(\boldsymbol t;\tau)$ in Eq.~\eqref{eq:essMHOM} for $\|\boldsymbol t\|=t=1$ and $\tau=0$ in the direction $\boldsymbol e=\boldsymbol t/t$.
	The particular nonclassical feature of this pure, two single-photon state~\eqref{eq:HOMstate} is highlighted by the node in the $z$ direction.
	In contrast, the MGF for a pure, classical coherent state in Eq.~\eqref{eq:essMcoherent} describes the shape of an ellipsoid, not a torus.
	The node relates to the truncated photon distribution of this state which is a signature of its nonclassicality and which has been discussed in the context of Eq.~\eqref{eq:node}.

	More generally, we get from Eq.~\eqref{eq:sampleMess} that $M_{\rm ess}(\boldsymbol e;0)=0$ if $p(n_a,0;\boldsymbol e)=0$ for all $n_a$.
	This means that a node will occur, $M_{\rm ess}(\boldsymbol e;0)\boldsymbol e=(0,0,0)^{\rm T}$, if the probability to have no photons in the second mode vanishes -- as $\sum_{n_a=0}^\infty p(n_a,0;\boldsymbol e)=0$ in that scenario.
	This is true for the considered state $|\psi_{\boldsymbol e}\rangle$ in Eq.~\eqref{eq:HOMstate} in the case $T=0$ or $R=0$, i.e., $\boldsymbol e=(0,0,\pm1)^{\rm T}$, as shown in Fig.~\ref{fig:HOM1}.

\begin{figure}[t]
	\includegraphics[width=6.5cm]{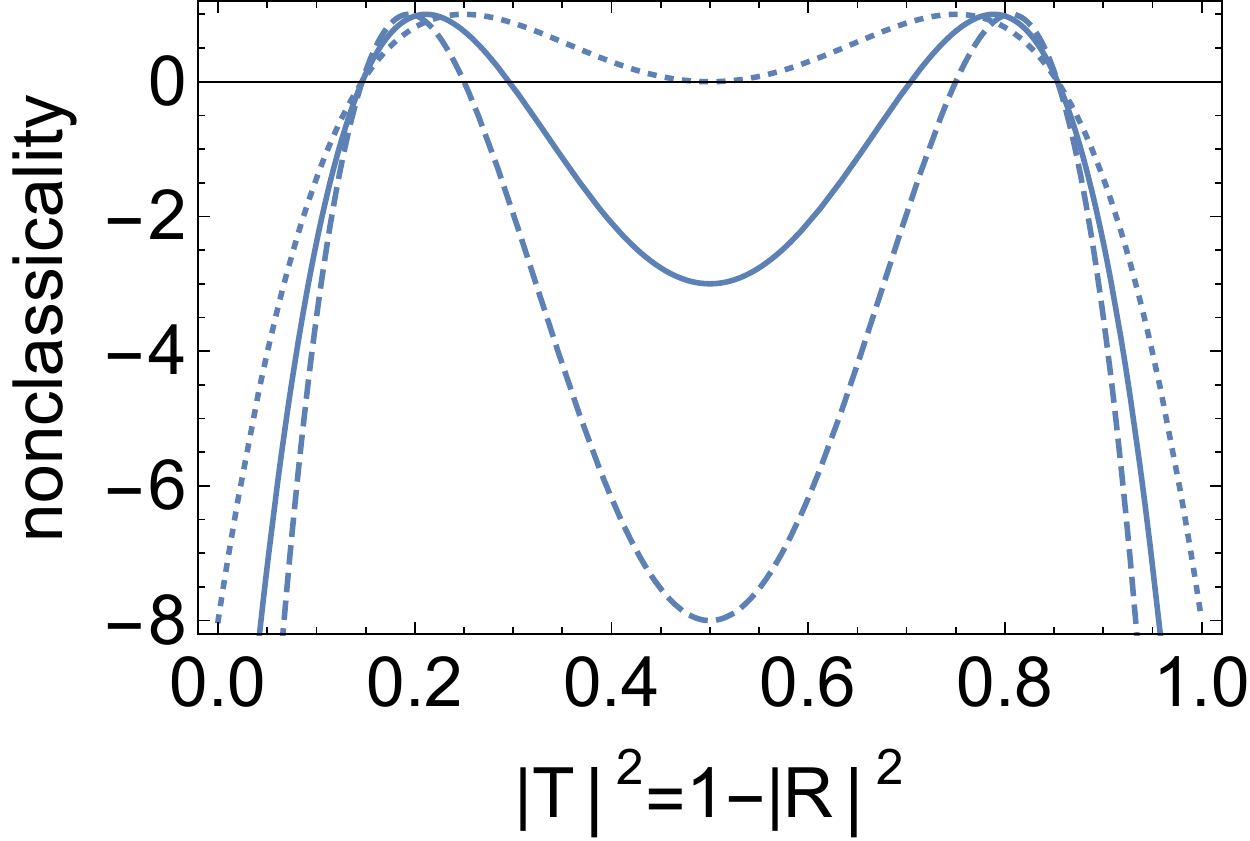}
	\caption{(Color online)
		The determinant in Eq.~\eqref{eq:second-order-NCL} is plotted for $\tau=\tau'=0=\boldsymbol t'$ as a function of $|T|^2$.
		The other vector is $\boldsymbol t=t\boldsymbol e$, with $\boldsymbol e$ as given in Eq.~\eqref{eq:Def-e-and-S}.
		A negative value of the determinant in the plot certifies essential nonclassical correlations.
		Note, there is no dependency on the phases of $T$ or $R$; cf. Eq.~\eqref{eq:essMHOM}.
		The parameter $t$ is chosen as $t=\sqrt{2}$ (dotted line), $t=\sqrt{3}$ (solid line), and $t=\sqrt{4}$ (dashed line).
		For $|T|^2=|R|^2=1/2$, we have a balanced beam splitter.
		The values $|T|^2=1$ or $|R|^2=1$ describe a perfect transmission or reflection, respectively.
	}\label{fig:HOM2}
\end{figure}

	In Fig.~\ref{fig:HOM2}, the verification of essential quantum correlations via the criterion~\eqref{eq:second-order-NCL} is shown for the state under study.
	It can be seen that the nonclassicality of the Hong-Ou-Mandel experiment can be clearly verified ($|T|\approx|R|$ yields a negative value) for amplitudes $\|\boldsymbol t\|>\sqrt 2$.
	Surprisingly, even the nonclassicality of the individual photons is demonstrated ($|T|\approx 1$ or $|R|\approx 1$ also yields a negativity) by the same criterion, even for a broader rage of amplitudes; see the dotted line for the case $\|\boldsymbol t\|=\sqrt 2$.

\section{Essential quantum correlations, polarization tomography, and imperfect click-counting detectors}\label{sec:4}
\begin{figure}[b]
	\includegraphics[width=7cm]{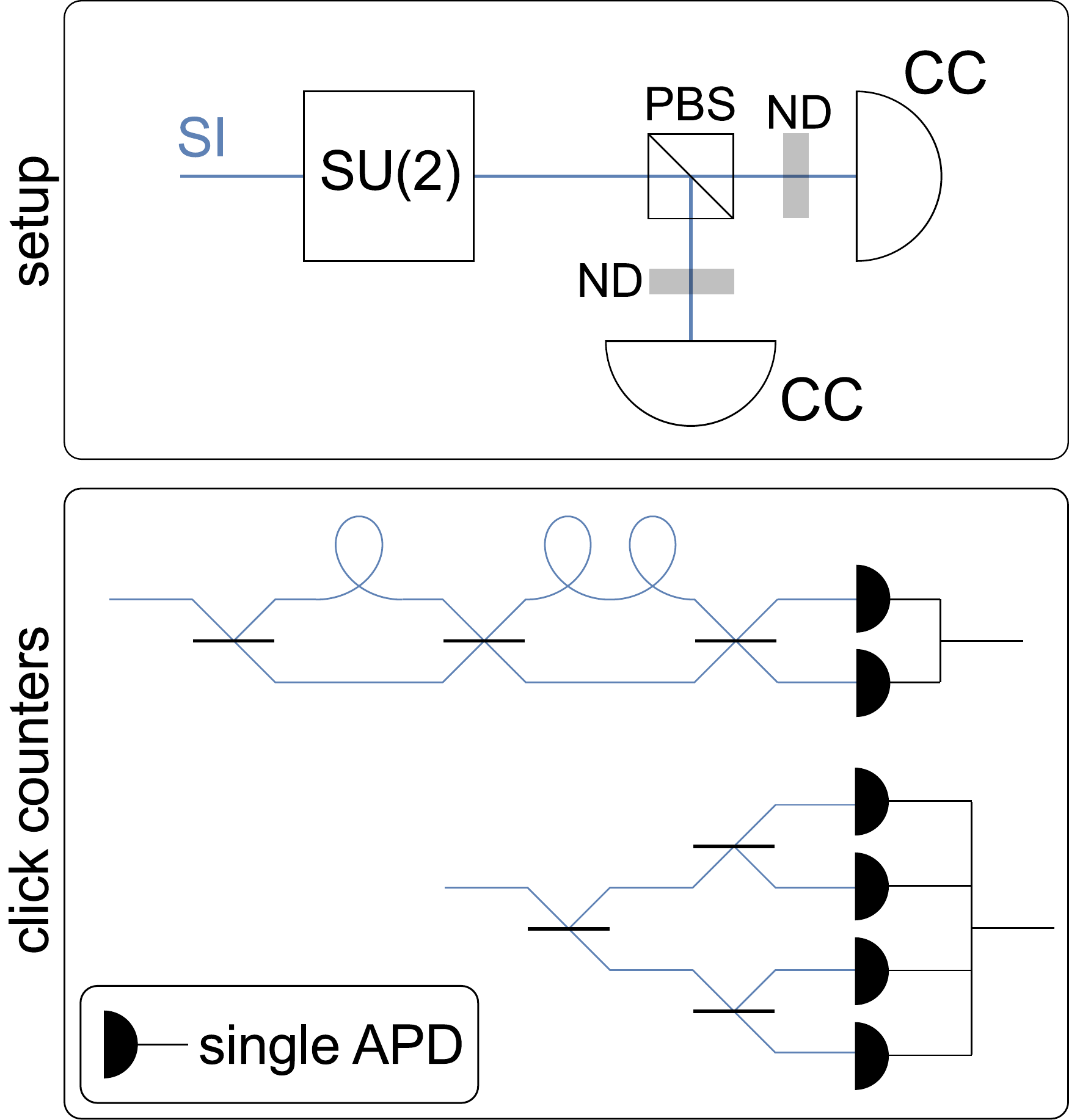}
	\caption{(Color online)
		Schematics of the setup for measuring the polarization of the signal (SI).
		The SU(2) transformation can be realized with half-wave and quarter-wave plates.
		PBS is a polarizing beam splitter, ND is a neutral density filter, and CC denotes a click-counting detector.
		Bottom pattern: Two possible implementations of a click-counter; see Ref.~\cite{S07} for a recent review.
		The upper version shows the time-bin~\cite{ASSBW03,FJPF03} and the lower version the spatial~\cite{PTKJ96,KB01} multiplexing configuration.
		In each step, the signal is split on a $50{:}50$ beam splitter.
		Avalanche photodiodes in the Geiger mode (APDs) serve as on-off detectors.
	}\label{fig:setup}
\end{figure}

	Let us now consider a realistic detection scenario, which allows one to apply our proposed method and which is outlined in Fig.~\ref{fig:setup}.
	The basic underlying scheme is a polarization tomography experiment, which has been intensively studied in theory and experiments in the single- and multiphoton domain~\cite{L05,WABGJJKMP05,L06,SZA10,MSPGRHKLMS12,SBKSL12,IARAS12,KRW14}.
	As it is advantageous for measuring $M_{\rm ess}$, we basically replace the standard photon counters with click-counting detectors in such layouts, which is described in the continuation of this section.

	We assume that the source in Fig.~\ref{fig:setup} emits a signal field of polarized light, where $\hat a$ and $\hat b$ describe the horizontal and vertical polarization modes, respectively.
	A combination of half- and quarter-wave plates allows one to implement arbitrary [${\rm SU}(2)$] transformations, $T=\cos(\vartheta/2)$ and $R=\sin(\vartheta/2)e^{-i\varphi}$~\cite{MJ66,AC09}.
	In this parametrization, the vector $\boldsymbol e$ takes the form
	\begin{align}\label{eq:SU2trafo}
		\boldsymbol e=\begin{pmatrix}\sin(\vartheta)\cos(\varphi)\\\sin(\vartheta)\sin(\varphi)\\\cos(\vartheta)\end{pmatrix}.
	\end{align}
	This renders it possible to scan all directions $\boldsymbol e$ of $M_{\rm ess}(t\boldsymbol e;\tau)$.
	After this manipulation, the two polarization components are spatially separated by a polarizing beam splitter (PBS).
	The outputs may be attenuated with neutral-density (ND) filters with intensity transmission efficiencies of $\varepsilon_a$ and $\varepsilon_b$, which serve later on as an additional degree of freedom.
	Until this point, one can see that a two-mode polarized, coherent input field, $|\alpha,\beta\rangle$, will result in the mean output photon numbers $\bar n_a$ and $\bar n_b$, which are
	\begin{align}
	\begin{aligned}
		\bar n_a=&\varepsilon_a|\cos(\vartheta/2)\alpha+\sin(\vartheta/2)e^{-i\varphi}\beta|^2	
		\\\text{and }
		\bar n_b=&\varepsilon_b|\cos(\vartheta/2)\beta-\sin(\vartheta/2)e^{i\varphi}\alpha|^2.
	\end{aligned}
	\end{align}
	Likewise, this may be also expressed via the Stokes parameters using Eq.~\eqref{eq:BosonStokesRelation}, $\bar n_{a(b)}=[\varepsilon_{a(b)}/2][\|\boldsymbol S\|\pm\boldsymbol e\cdot\boldsymbol S]$.

	Finally, let us assume that we have two click-counting (CC) detection systems in Fig.~\ref{fig:setup}.
	Each of them consists of $D_a$ and $D_b$ avalanche photodiodes (APDs) having quantum efficiencies $\eta_a$ and $\eta_b$ and dark-count rates $\nu_a$ and $\nu_b$, respectively.
	The joint click statistics, i.e., the probability that we have $i$ clicks from the first APD array simultaneous with $j$ clicks from the second detector array, can be written in form of a quantum version of the binomial statistics~\cite{SVA12,SVA13}:
	\begin{align}
		&c(i,j;\boldsymbol e)
		\\\nonumber=&\left\langle{:}
			\binom{D_a}{i}\hat m_a^{D_a-i}\left(\hat 1-\hat m_a\right)^{i}
			\binom{D_b}{j}\hat m_b^{D_b-j}\left(\hat 1-\hat m_b\right)^{j}
		{:}\right\rangle,
	\end{align}
	where $\hat m_a=\exp(-\varepsilon_a\eta_a\hat a_{\boldsymbol e}^\dag\hat a_{\boldsymbol e}/D_a-\nu_a)$ and correspondingly for $\hat m_b$, which depends on the vector $\boldsymbol e$ in Eq.~\eqref{eq:SU2trafo}.
	This class of detectors cannot truly resolve the number of photons, as, first, it is limited by a finite quantum efficiency and dark-count rate and, second, it can only deliver a finite number of clicks albeit the photon statistics is an infinite one.
	However, based on such detection systems, one can nevertheless infer quantum properties of light, e.g., sub-binomial light~\cite{SVA12a,BDJDBW13,HSPGHNVS16}. 

\begin{figure*}
	\includegraphics[width=5.5cm]{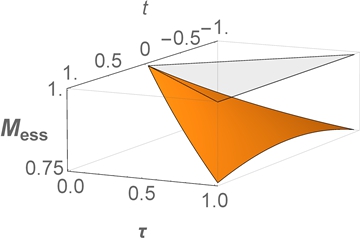}\hspace*{0.5cm}
	\includegraphics[width=5cm]{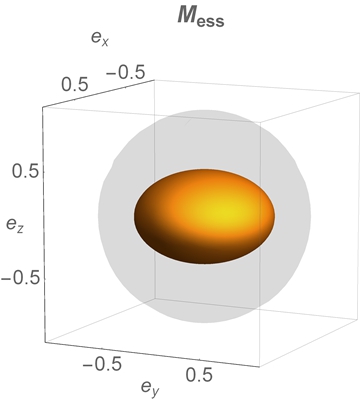}\hspace*{0.5cm}
	\includegraphics[width=5.5cm]{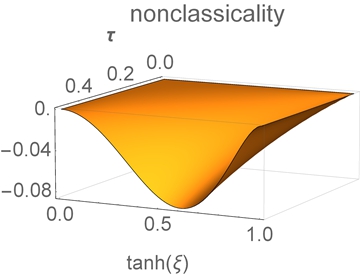}
	\caption{(Color online)
		Left:
		The measurable MGF, $M_{\rm ess}(t\boldsymbol e;\tau)$, is shown depending on $\tau$ and $-\tau\leq t\leq\tau$, with $\boldsymbol e=(0,0,1)^{\rm T}$ and $\tanh(\xi)=1/2$.
		The transparent gray area shows the same for a vacuum state for comparison.
		Center:
		$M_{\rm ess}(t\boldsymbol e;\tau)$ in the direction of $\boldsymbol e$ is depicted as a function of $\boldsymbol e=(\sin(\vartheta)\cos(\varphi),\sin(\vartheta)\sin(\varphi),\cos(\vartheta))^{\rm T}$, with $\tau=t=1$ and $\tanh(\xi)=3/4$.
		The transparent surface shows the result for a vacuum state.
		Right: 
		The verification of nonclassicality (negative values) by the determinant in Eq.~\eqref{eq:second-order-NCL} is given as a function of $\tanh(\xi)$ and $\tau$ ($0\leq 2\tau\leq1$), with $\boldsymbol e=\boldsymbol e'=(0,0,1)^{\rm T}$ and $\tau'=t'=\tau=-t$.
	}\label{fig:ex2}
\end{figure*}

	From the click statistics, we can directly sample the $(k,l)$th normally ordered moment~\cite{SVA13,SBVHBAS15},
	\begin{align}
	\begin{aligned}
		&\mu_{k,l}=\left\langle{:}
			\hat m_a^{k}\hat m_b^{l}
		{:}\right\rangle	
		=\sum_{i,j=0}^{D_a,D_b}\frac{\binom{D_a-i}{k}\binom{D_b-j}{l}}{\binom{D_a}{k}\binom{D_b}{l}}c(i,j;\boldsymbol e)
		\\=&e^{-\nu_a k-\nu_b l}\left\langle{:}\exp\left[
			-\frac{\eta_a\varepsilon_a}{D_a}\hat a_{\boldsymbol e}^\dag\hat a_{\boldsymbol e}
			-\frac{\eta_b\varepsilon_b}{D_b}\hat b_{\boldsymbol e}^\dag\hat b_{\boldsymbol e}
		\right]{:}\right\rangle,
	\end{aligned}
	\end{align}
	for $0\leq k\leq D_a$ and $0\leq l\leq D_b$ and defining $\binom{x}{y}=0$ for $y>x$.
	The dark-count rates can be measured by blocking the signal, as we have in this case: $\mu_{1,0}=e^{-\nu_a}$ and $\mu_{0,1}=e^{-\nu_b}$.
	They contribute to the moments $\mu_{k,l}$ as factors and, thus, one can simply divide by $e^{-k\nu_a-l\nu_b}$ to delete these contributions.
	This means, we can also directly assume that $\nu_a=\nu_b=0$.

	Now, let us insert Eq.~\eqref{eq:output-numbers}, which yields
	\begin{align}\label{eq:momident1}
	\begin{aligned}
		\mu_{k,l}=
		&\left\langle{:}
			\exp\left[-\left(\frac{k\varepsilon_a\eta_a}{2D_a}+\frac{l\eta_b\varepsilon_b}{2D_b}\right)\hat N\right.
		\right.
		\\
		&\phantom{\langle{:}\exp[{-}}\left.
			\left.+\left(\frac{l\eta_b\varepsilon_b}{2D_b}-\frac{k\eta_a\varepsilon_a}{2D_a}\right)\boldsymbol e\cdot\boldsymbol{\hat S}\right]
		{:}\right\rangle.
	\end{aligned}
	\end{align}
	We can identify $t$ and $\tau$ by comparing with Eq.~\eqref{eq:expectMGF0}, 
	\begin{align}\label{eq:momident2}
		\tau=\frac{k\varepsilon_a\eta_a}{2D_a}+\frac{l\eta_b\varepsilon_b}{2D_b}
		\text{ and }
		t=\frac{l\eta_b\varepsilon_b}{2D_b}-\frac{k\eta_a\varepsilon_a}{2D_a}.
	\end{align}
	The measurable parameter range of $t$ and $\tau$ can be deduced as follows.

	From $k=D_a$ and $l=D_b$, we can see that $\tau$ is bounded as $0\leq\tau\leq(\eta_a+\eta_b)/2\leq1$, where $0\leq \varepsilon_a,\varepsilon_b\leq 1$ was used.
	Without the neutral-density filters ($\varepsilon_a=\varepsilon_b=1$), only discrete values for the different $k$ and $l$ combinations could be measured.
	Additionally, we get for $k=D_a$ and $l=0$ or $k=0$ and $l=D_b$ that $-1\leq-\eta_a\leq t\leq \eta_b\leq 1$.
	This means that we can access $M_{\rm ess}(t\boldsymbol e;\tau)$ in intervals that are included in $[-1,1]$ for $t$ and $[0,1]$ for $\tau$.
	Moreover, the absolute value of $t$ is always less than or equal to $\tau$, $|t|\leq\tau$.
	Let us stress that we have a restricted parameter range, but we can access the MGF in this domain directly -- by applying Eqs.~\eqref{eq:momident1} and~\eqref{eq:momident2} -- even though we have no photon-number resolution and finite quantum efficiencies.
	This also means that the nonclassicality criteria following from inequality~\eqref{eq:genMessMOM} can be easily applied.

	We may study a type II parametric down-conversion process, which is described by the Hamiltonian $\hat H\propto \hat a^\dag\hat b^\dag+\hat a\hat b$.
	The generated photon pairs are produced in the two polarization modes and they can propagate colinearly~\cite{T01,KOW01,K03,MLSWUSW08}.
	The resulting state is a two-mode squeezed-vacuum state,
	\begin{align}
		|\xi\rangle=\frac{1}{\cosh(\xi)}\exp\left[-\tanh(\xi)\hat a^\dag\hat b^\dag\right]|{\rm vac}\rangle,
	\end{align}
	which is parametrized by the squeezing parameter $\xi\geq0$.
	Properties of such polarization squeezed and entangled states have been experimentally studied~\cite{KLLRS02,LLA09}.
	For $\xi=0$ [$\tanh(\xi)=0$], we have the classical vacuum state, i.e., a coherent state with a vanishing coherent amplitude, which exhibits no quantum correlations.
	Conversely, the case $\xi\to\infty$ [$\tanh(\xi)=1$] yields an Einstein-Podolsky-Rosen entangled state.
	Again, the integral~\eqref{eq:essMHusimi} with the Husimi function~\eqref{eq:Husimi} can be computed, which gives the desired result.
	It reads
	\begin{align}
	\begin{aligned}
		&M_{\rm ess}(t\boldsymbol e;\tau)=
		\langle\xi|{:}\exp\left[-\lambda_a\hat a_{\boldsymbol e}^\dag\hat a_{\boldsymbol e}-\lambda_b\hat b_{\boldsymbol e}^\dag\hat b_{\boldsymbol e}\right]{:}|\xi\rangle	
		\\=&\Big[\big[\cosh^2(\xi)-(1-\lambda_a)(1-\lambda_b)\sinh^2(\xi)\big]^2
		\\&-\sin^2(\vartheta)\sinh^2(\xi)\cosh^2(\xi)(\lambda_a-\lambda_b)^2\Big]^{-1/2},
	\end{aligned}
	\end{align}
	for $0\leq\lambda_a,\lambda_b\leq 1$, $0\leq\vartheta\leq \pi$, and $\xi\geq0$.
	Let us also point out that $\tau=\lambda_a+\lambda_b$ and $t=\lambda_b-\lambda_a$ [Eq.~\eqref{eq:lambdas}].

	Based on these formulas, we can now investigate the state of light and the optical measurement scheme under study with our techniques; see Fig.~\ref{fig:ex2}.
	The left plot in Fig.~\ref{fig:ex2} shows $M_{\rm ess}(t\boldsymbol e;\tau)$ for the parameter range $\tau$ and $t$ which is accessible with this kind of polarization tomography.
	The measurement direction $\boldsymbol e$ is fixed.
	We observe a decaying behavior for the examples of a two-mode squeezed-vacuum state ($\xi>0$) in contrast to the constant value for a vacuum state (transparent area, $\xi=0$).

	In the center plot of Fig.~\ref{fig:ex2}, we fixed $\tau$ and the amplitude of $\boldsymbol t$ and considered all measurement directions and depicted the vector $M_{\rm ess}(t\boldsymbol e;\tau)\boldsymbol e$.
	A sphere represents the result for a vacuum state.
	The properties of the pure squeezed-vacuum state are visualized by the fact that the squeezed state yields an ellipsoid with two major principle ($x$ and $y$ direction) axes and only one minor principle axis ($z$ direction) in this representation, which is not possible for classical coherent states; cf. the discussion below Eq.~\eqref{eq:essMcoherent}.

	In the right plot of Fig.~\ref{fig:ex2}, we see the successful verification of nonclassicality via the directly measured MGF as a function of $\tau>0$ and $\xi>0$, i.e., $\tanh(\xi)\in]0,1[$.
	We consider the simplest case of a full transmission, $T=1$ [$\boldsymbol e=(0,0,1)^{\rm T}$].
	For all non-trivial parameters, the second-order determinant in Eq.~\eqref{eq:second-order-NCL} identifies the nonclassicality of the generated polarization state of light, as it is always negative.
	This holds true even for arbitrarily small efficiencies ($\tau$ close to zero) and small squeezing values.
	At first sight, the values of the negativities seem to be small.
	However, it has been experimentally demonstrated (e.g., in Ref.~\cite{SBVHBAS15}) that those negativities are still detectable with a high statistical significance.
	Moreover, let us point out that the nonclassicality goes to zero for $\xi\to\infty$ due to the saturation of the detection system.
	That is, the intensity is so high that all APDs click all the time, which is also achievable with a classical coherent signal with a large coherent amplitude.


\section{Conclusions}\label{sec:5}
	In summary, we have formulated an operational approach to uncover the accessible quantum correlations in optical systems.
	To do so, we introduced the notion of the essential quantum correlations and the corresponding phase-space distribution was derived.
	In particular, we focused on a four-port interferometer and a subsequent measurement of the output intensities for justifying our operational definition.
	In the following step, a reconstruction technique was formulated in terms of the photon-number distribution and a generalized moment-generating function.
	Based on the latter function, we also derived necessary and sufficient criteria for the essential nonclassicality and, in particular, studied the relation to other nonclassicality tests based on the second order of our general set of inequalities.
	The essential nonclassicality of the Hong-Ou-Mandel interference experiment was verified.
	Finally, we considered a realistic measurement scheme and showed that our approach is applicable even in the presence of losses and without photon-number resolution.
	It was based on click-counting measurements whose moments directly describe the desired moment-generating function for a certain parameter range.

	Let us emphasize some further aspects of our treatment.
	The main aim of this work was the operational definition of nonclassicality which is truly experimentally available with a certain setup.
	With such a practicable definition, we demonstrated that it is feasible to uncover phase-sensitive aspects of the nonclassicality of light fields without relying on the generation of proper reference beams.
	We placed minimal assumptions onto experimental implementations to ensure, to some extend, the general validity of our findings.
	Along with our rigorous theoretical handling, we also studied the impact of imperfections which are unavoidable in any realistic experiment.

	The presented methods may also be the basis for further considerations of the essential quantum correlations in more general scenarios.
	In the here-presented first step, our focus was on particular measurement layouts and on the nonclassicality of two correlated optical modes.
	For future studies and following the here-introduced steps, our findings could be extended to other scenarios, e.g., to capture temporal correlations in optical fields, to include other notions of nonclassical multimode correlations, such as entanglement, or to study other systems, e.g., atomic ensembles.


\section*{Acknowledgments}
	J.S. and .W.V have received funding from the European Union's Horizon 2020 research and innovation program under grant agreement No 665148.
	J.S. gratefully acknowledges financial support from the Oklahoma State University where part of this work was done.


\end{document}